\documentclass[epj,referee]{svjour} 

\usepackage{graphics}

\newcommand{\prt}{\partial} 
\newcommand{\prm}{\prime}

\begin{document}

\title{Hydraulic Jump in One-dimensional Flow}

\subtitle{}

\author{Subhendu B. Singha\inst{1}\thanks{\tt{sbsingha@theory.tifr.res.in}},  
Jayanta K. Bhattacharjee\inst{2}\thanks{\tt{tpjkb@mahendra.iacs.res.in}} 
\and Arnab K. Ray\inst{3}\thanks{\tt{arnab@mri.ernet.in}}}

\institute{Department of Theoretical Physics, Tata Institute of
Fundamental Research, Homi Bhaba Road, Mumbai 400005, India 
\and
Department of Theoretical Physics, Indian Association for the
Cultivation of Science, Jadavpur, Kolkata 700032, India 
\and
Harish--Chandra Research Institute, Chhatnag Road, Jhunsi, Allahabad
211019, India}

\date{Received: date / Revised version: date}

\abstract{In the presence of viscosity the hydraulic jump in one
dimension is seen to be a first-order transition. A scaling relation
for the position of the jump has been determined by applying an
averaging technique on the stationary hydrodynamic equations. This
gives a linear height profile before the jump, as well as a clear
dependence of the magnitude of the jump on the outer boundary
condition. The importance of viscosity in the jump formation has been 
convincingly established, and its physical basis has been understood 
by a time-dependent analysis of the flow equations. In doing so, a 
very close correspondence has been revealed between a perturbation 
equation for the flow rate and the metric of an acoustic white hole. 
We finally provide experimental support for our heuristically 
developed theory.
}

\maketitle

\section{Introduction}
\label{sec1}

A stream of water impinging vertically on to a horizontal surface
spreads out radially in a thin sheet along the plane from the point
of impingement, and at a certain radius the height of the flowing 
layer of water suddenly increases. In this two-dimensional flow, such 
an abrupt increase in the level of the liquid is known as the circular 
hydraulic jump \cite{ot66,tan49,wat64}. It is a familiar observation, 
seen everyday in the kitchen sink. A similarly abrupt increase in 
the height --- a jump --- occurs in the one-dimensional flow as 
well, and this phenomenon finds mention in many introductory text books 
on hydrodynamics \cite{wj79,ll87,gra87,fab95,ghpm01}. A very regularly
cited practical example of the jump in the one-dimensional flow is the 
passage of a tidal bore up a river \cite{fab95}. However, the texts 
include viscosity --- arguably the primary physical cause of the jump
--- only through a phenomenologically added friction term \cite{gra87}. 
Consequently, starting from the Navier-Stokes equation, it has not been
possible to predict the position of the jump in terms of the volumetric 
flow rate. For the two-dimensional flow, on the other hand, the role of 
viscosity has been very clearly taken into account in the works of Bohr 
et al. \cite{bdp93,behh96,bpw97}. This has led to a scaling dependence for
the position of the jump on the volumetric flow rate. The two-dimensional 
problem, however, is sufficiently complicated, and in predicting the 
position of the jump in this case, knowledge of 
asymptotic solutions has been necessary to a fair extent.
 
Motivated by the methods applied to study the two-dimensional flow,
and by the results obtained thereof, we make a similar analysis of
the one-dimensional flow in this work. We derive a profile of the
height of the liquid layer in a one-dimensional open-channel flow by
making transparent approximations about the nature of the flow. We
find that the position of the jump is sensitive to the vertical
profile of the velocity field, and we make use of this dependence to
conclude that the profile is far from parabolic and resembles more
closely a turbulent profile. There exists no definite result for the
magnitude of the jump. Text books \cite{ll87,gra87,fab95,ghpm01}
analyse the problem in one dimension, from which it can be shown that
for $h_1$ and $h_2$ being the heights of the liquid layer before and
after the jump respectively, the ratio $h_2/h_1$ is unity for the
critical value of the Froude number $\mathcal{F}$ 
(i.e for $\mathcal{F} = 1$), and increases smoothly as the Froude 
number is increased from unity. This indicates that the jump is a 
second-order transition. Contradicting this viewpoint, in our present 
analysis we put forward a heuristic picture of the role of 
viscosity and find that in its presence the jump attains a finite 
value at the critical Froude number. This is exactly what happens 
in a first-order transition. In our calculations we also establish 
a connection between the magnitude of the jump and the outer boundary 
condition. We use steady hydrodynamic equations to obtain the position 
and the height of the jump. In the process we reinforce our conclusion 
that the jump is of the nature of a first-order transition. 

Furthermore, we carry out a time-dependent linearised perturbative 
analysis about the steady flow solution. We perturb the steady and 
constant flow rate, and see that viscous dissipation, of course, drives 
the system back towards stability, but what we have made a note of with 
much greater interest is that the equation for perturbation in the flow 
has a remarkable degree of closeness with a metric that implies an acoustic
white hole. This analogy, coupled with some characteristic time scales
obtained through the perturbative analysis, enables us to argue for a 
physical basis behind the formation of the jump. 

It is in a largely heuristic spirit that we have made our 
theoretical foray into the channel flow problem. To bolster our 
arguments, we have therefore brought forth some experimental 
evidence in support of our theory. We have measured the magnitude
of the jump and the height profile of the flow before the jump. 
On both counts we find good agreement with our theory. From our 
experimental data we can also easily infer that the parabolic 
profile in the vertical direction has no validity. 

\section{Role of Viscosity : A Heuristic Study}
\label{sec2}

\begin{figure}
\begin{center}
\resizebox{0.9 \hsize}{!}{\includegraphics{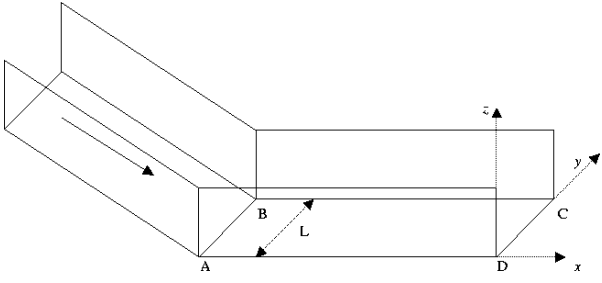}}
\end{center}
\caption{The geometry of the experimental set up. The horizontal flow
starts from the edge $AB$.}
\label{Fig1}
\end{figure}

The flow occurs in a channel of width $L$, with $L$ being very much
greater than the depth of the liquid layer, as has been schematically
shown in Fig.\ref{Fig1}. At some point the depth changes from $h_1$ to
$h_2$. We work with a control volume which extends from a point before
the jump to a point after the occurrence of the jump. Ignoring the
effect of viscosity, the continuity equation can be written down as
\begin{equation}
\label{e1}
Lh_1u_1=Lh_2u_2=Q
\end{equation}
in which $u_1$ and $u_2$ are the flow velocities before and after the
jump respectively, while $Q$ is the volumetric flow rate. The momentum
change $\rho Q \left (u_2-u_1 \right)$ per unit time is brought about 
by the force due to the pressure difference. This gives the balancing 
condition
\begin{equation}
\label{e2}
\frac{1}{2}\rho g L \left(h_1^2 - h_2^2 \right)  = \rho Q
\left(u_2-u_1 \right)
\end{equation}
First writing $\mathcal{H}=h_2/h_1$ and the Froude number
$\mathcal{F}$ as $\mathcal{F}=u_1^2/gh_1$, we combine Eqs.(\ref{e1})
and (\ref{e2}), to get
\begin{equation}
\label{e3}
\mathcal{H}\left(1+ \mathcal{H} \right)-2 \mathcal{F}=0
\end{equation}
which can be solved to get $2 \mathcal{H} =-1+\sqrt{1+ 8
\mathcal{F}}$. If we now write $\mathcal{F} = 1+ \vartheta$ with 
the condition $0< \vartheta \ll 1$, then to first order 
in $\vartheta$ we will get $\mathcal{H} =1+2\vartheta/3$, which 
establishes the standard text book 
interpretation \cite{ll87,gra87,fab95} of the jump being a 
continuous transition as a function of $\mathcal{F}$.

\begin{figure}
\begin{center}
\resizebox{1.0 \hsize}{!}{\includegraphics{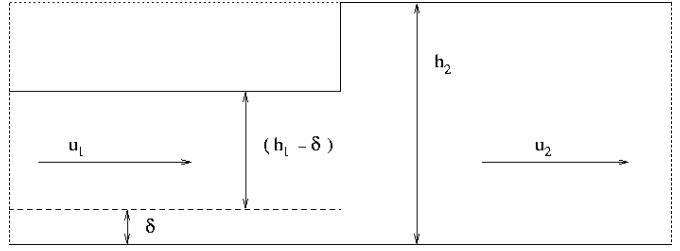}}
\end{center}
\caption{Control volume of the flow, demarcated by the dotted lines.
The boundary layer has an average thickness of $\delta$.}
\label{Fig2}
\end{figure}

We need to modify the above picture in the presence of viscosity. The
most important contribution of viscosity will be the formation of a
boundary layer. Practically speaking, the thickness of the boundary 
layer increases as the flow progresses along the plane, but in our
control volume we have constrained the flow to have an average 
thickness of $\delta$, within which the velocity increases from zero to 
$u_1$, while over the depth of $h_1-\delta$, the flow velocity remains 
at a constant value of $u_1$. This state of affairs has been 
schematically represented
in Fig.\ref{Fig2}. After the jump, beyond a mixing zone 
characterized by vortices, the flow has a mean speed $u_2$ and an increased 
depth $h_2$. However, it must be noted here that following the jump, the
flow has been known to be turbulent --- something that may very readily 
be appreciated from the analogous case of the circular hydraulic 
jump \cite{hans97} --- and it becomes too much complicated to be thought 
of simply in terms of a boundary layer. Therefore we keep our analysis 
of the control volume confined to the flow region before the jump. 
Within the boundary layer we make use of an arbitrary profile i.e. 
$u(z)=u_1 \varphi (z/\delta)$
with $ \varphi (\xi) \leq 1$ for all $\xi$, and with $\xi$
itself constrained by the range $0 \leq \xi \leq 1$. The continuity
equation now reads
\begin{equation}
\label{e4}
Lu_1 \left( h_1-\delta \right) + L\int^{\delta}_{0}u_1  \varphi
\bigg(\frac{z}{\delta} \bigg) \, \mathrm{d}z = Lu_2h_2 = Q
\end{equation}
which, following some manipulations, can be rendered as
\begin{equation}
\label{e5}
L u_1 \left[h_1-\delta \left(1- I_1 \right) \right] = L u_2 h_2 = Q
\end{equation}
where $I_1=\int_0^1 \varphi (\xi) \, \mathrm{d}\xi$. In the presence
of viscosity, this is the modified form of Eq.(\ref{e1}).

The force balance equation requires a similar modification in the
momentum flow rate. Before the jump the amount of momentum entering
the control volume per unit time is
\begin{eqnarray}
\label{e6}
L\rho\left(h_1-\delta \right)u_1^2 &+& L\rho \int^{\delta}_{0} u_1^2
\varphi^2 \bigg(\frac{z}{\delta} \bigg)\, \mathrm{d}z \nonumber \\ & &
= \rho Qu_1 - L\rho u_1^2\delta \left(I_1-I_2 \right)
\end{eqnarray}
where $I_2=\int_0^1 \varphi^2(\xi)\, \mathrm{d}\xi$. The amount of
momentum leaving the control volume per unit time is $\rho Qu_2$ and
the force due to pressure on the control volume is $\rho
Lg\left(h_1^2 - h_2^2 \right)/2$ acting to the right. The force
balance equation now becomes,
\begin{equation}
\label{e7}
\rho Q\left(u_2-u_1 \right) + L\rho\delta u_1^2 \left(I_1-I_2 \right)
= \frac{1}{2}\rho gL\left( h_1^2 - h_2^2 \right)
\end{equation}
Using the continuity condition as expressed in Eq.(\ref{e5}), we find
\begin{eqnarray}
\label{e8}
u_1^2 \big[h_1 &-& \delta (1 - I_1) \big] \left[ \frac{h_1}{h_2}-1 -
\frac{\delta \left(1-I_1 \right)}{h_2} \right] \nonumber \\ &+& \delta
u_1^2 \left( I_1-I_2 \right) =\frac{g}{2}\left(h_1^2-h_2^2 \right)
\end{eqnarray}
As we have done for the inviscid case, we use the same definition of
$\mathcal{H}$ and $\mathcal{F}$, and obtain an expression from 
Eq.(\ref{e8}) that reads as 
\begin{eqnarray}
\label{e9}
2 \mathcal{F} \left(\frac{1}{\mathcal{H}} - 1 \right) &-& 2\mathcal{F}
\frac{\delta}{h_1} \left[ 2 \left(1-I_1 \right)\frac{1}{\mathcal{H}}
+I_2-1 \right] \nonumber \\ & & +2 \mathcal{F} \frac{\delta^2}
{h_1^2}\left(1-I_1 \right)^2\frac{1}{\mathcal{H}} = 1 - \mathcal{H}^2
\end{eqnarray}
from which we can easily see that if $\delta = 0$, i.e. if the effect
of viscosity is neglected, then the result given by Eq.(\ref{e3}) will
be recovered. In this inviscid limit, prescribing $\mathcal{F}=1+
\vartheta$ leads to $\mathcal{H} =1+\varepsilon$ with $\varepsilon =
2\vartheta /3$. In the presence of viscosity, we once again seek a
jump solution by writing $\mathcal{H} =1+\varepsilon$ with $\varepsilon >
0$ for $\mathcal{F}=1+ \vartheta$. In the limit 
$\vartheta \longrightarrow 0$, we will then have the cubic equation
\begin{eqnarray}
\label{e10}
\varepsilon^3 + 3\varepsilon^2 + 2\frac{\delta}{h_1}\left(1-I_2 \right)
\bigg[\varepsilon &-& \frac{\left(1+I_2-2I_1 \right)}{\left(1-I_2
\right)} \bigg] \nonumber \\ 
&+& 2 \frac{\delta^2}{h_1^2} \left(1-I_1 \right)^2 = 0
\end{eqnarray}
It is clear that there can be no positive root of $\varepsilon$ which is
greater than $(1+I_2-2I_1)/(1-I_2)$ since in that case all terms on
the left hand side of Eq.(\ref{e10}) will be positive. For a Couette
profile $I_1=1/2$ and $I_2=1/3$. This gives $\varepsilon=0.5$ as an
upper limit. For the parabolic profile $I_1=2/3$ and $I_2=8/15$,
giving $\varepsilon=0.43$ as a maximum value. On the other hand, a 
continuous transition would imply that $\varepsilon =0$. If we use 
this value of $\varepsilon$ in Eq.(\ref{e10}), we will get the 
condition $\delta/h_1 = (1+I_2-2I_1)/(1-I_1)^2$, which, for both
the Couette and the parabolic profiles should give a value of 
$\delta/h_1$ to be greater than unity. This is physically an
untenable result, because $\delta$ is the average thickness of the
boundary layer taken in the region before the jump, and as such, its
value must be less than $h_1$. This inconsistency is indication 
enough that $\varepsilon$ is not zero, and therefore the transition 
is not continuous. 

For a more methodical evaluation of the roots of $\varepsilon$, we will
have to solve Eq.(\ref{e10}). To that end, for notational convenience
we write the third degree expression on the left-hand side of 
Eq.(\ref{e10}) as $\Phi(\varepsilon)$, and then solve for 
$\Phi(\varepsilon)=0$. To find the roots of this cubic equation, 
it would be necessary to eliminate the second degree term in $\varepsilon$,
by the suitable substitution $\varepsilon = \zeta -1$. This will then
render Eq.(\ref{e10}) in the standard form as $\zeta^3 + \mathcal{P}
\zeta + \mathcal{Q}=0$, where $\mathcal{P} = -3 + 2(1-I_2)\delta/h_1$
and $\mathcal{Q} = 2[1 - (1-I_2)\delta/h_1 + (1-I_1)^2 (\delta/h_1)^2]$.
The discriminant, $\mathcal{D}$, is given by 
$\mathcal{D} = (\mathcal{Q}^2/4) + (\mathcal{P}^3/27)$. 

\begin{figure}
\begin{center}
\resizebox{1.0 \hsize}{!}{\includegraphics{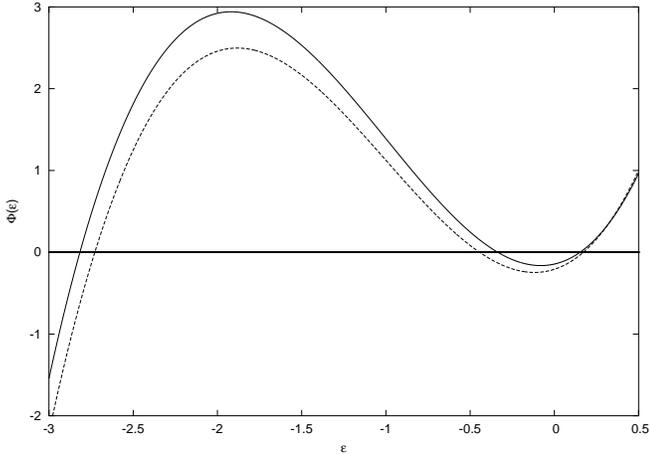}}
\end{center}
\caption{Plot of $\Phi(\varepsilon)$ versus $\varepsilon$. The roots
of $\varepsilon$ as given by Eq.(\ref{e10}), are to be found for
$\Phi(\varepsilon) = 0$. Both plots
have been drawn for $\delta/h_1 = 0.5$, with the continuous curve
representing the parabolic profile, and the dotted curve representing
the profile for the Couette flow.}
\label{Fig3}
\end{figure}

We choose $0.5$ as a fiduciary value for $\delta/h_1$, and using this
number for both the parabolic profile and the profile for the Couette flow, 
we find $\mathcal{D} < 0$. This must then imply that there should be three 
real roots of $\varepsilon$ for $\Phi(\varepsilon)=0$. This is very much 
in keeping with the fact that the function $\Phi(\varepsilon)$ has two 
real turning points, which can be obtained from the condition 
$\Phi^{\prime}(\varepsilon)=0$. These points are at $\varepsilon = -1 
\pm \sqrt{1 - (2/3)(1-I_2)\delta/h_1}$. Further, when $\varepsilon= 0$, the
function $\Phi(\varepsilon)$ has a negative value. This, in conjunction
with the fact that $\Phi(\varepsilon) \longrightarrow + \infty$ when 
$\varepsilon \longrightarrow + \infty$, could only mean that at most there 
should be only one positive real root of $\varepsilon$, a conclusion that 
has been clearly illustrated in Fig.\ref{Fig3}, in which we have plotted 
$\Phi (\varepsilon)$ against $\varepsilon$. For both the Couette and 
the parabolic profiles, three roots are to be found for 
$\Phi(\varepsilon)=0$, and in each case only one of these roots is positive. 
A numerical evaluation by the bisection method shows that for a parabolic 
profile the positive root of $\varepsilon$ is about $0.15$, while for 
the Couette flow it is $0.17$. For both these cases we have also assured 
ourselves that even for the limiting case of $\delta/h_1$ being unity, 
$\epsilon$ will have a positive root. The inescapable conclusion 
therefore is that the jump is a first-order transition, independent of 
the vertical velocity profile, because at the critical Froude number 
($\mathcal{F}=1$) the jump has a finite non-zero value. The present 
analysis drives home the point that when dissipation is included in 
the conventional control volume analysis, the second-order transition
immediately changes to a first-order transition.

\section{A Hydrodynamic Analysis from the Steady Flow Equations}
\label{sec3}

We consider an open rectangular channel of width $L$ (as shown
schematically in Fig.\ref{Fig1}) in which a liquid is flowing in
streamline motion. An arrangement is made such that the liquid flows
down an inclined channel and starts its one dimensional motion in the
horizontal plane from the edge $AB$. The $x$ axis is chosen in the
direction of flow, the $y$ axis along the width of the channel and the
$z$ axis in the vertical upward direction. At any point, the velocity
of the liquid is, in general, a function of the coordinates $x$, $y$
and $z$. In our present arrangement the width of the channel is
sufficiently large (about $9.1 \, \mathrm{cm}$) when compared with
the height of the liquid layer (which is of the order of a few
millimetres), and we can therefore assume that there is no variation
along the $y$ axis for the velocity. On the other hand, both its $x$
component, $u$, and $z$ component, $w$, would have a spatial
dependence in the form $u \equiv u(x,z)$ and $w \equiv w(x,z)$.

For an incompressible fluid, the local continuity equation gives,
\begin{equation}
\label{e11}
\frac{\prt u}{\prt x}  + \frac{\prt w}{\prt z} = 0
\end{equation}
and for a not very viscous liquid (e.g. water) the Navier-Stokes
equation in the boundary-layer approximation \cite{sch68} gives
\begin{equation}
\label{e12}
u \frac{\prt u}{\prt x}  + w \frac{\prt u}{\prt z} = -g
\frac{\mathrm{d}h} {\mathrm{d}x}  +\nu \frac{\prt^2u}{\prt z^2}
\end{equation}
where $h \equiv h(x)$ is the height of the liquid layer at a distance
$x$, and $\nu$ is the kinematic viscosity. The boundary conditions of
the flow are $u(x,0)=w(x,0)=0$, and $\prt u/\prt z =0$ at $z=h(x)$. 
In addition to this, the condition for constant volume flux gives
\begin{equation}
\label{e13}
L\int_0^{h(x)}u(x,z) \, \mathrm{d}z=Q
\end{equation}
We have assumed here that the shearing stress is zero at the free 
surface $z=h(x)$, since the viscosity of air is negligible. In the
boundary-layer approximation it is assumed that the vertical velocity
$w(x,z)$ is very small compared with the horizontal component
$u(x,z)$. Furthermore, the variation of $u(x,z)$ is much faster in
the $z$ direction as compared with the $x$ direction.

At this stage we follow Bohr et al. \cite{bdp93} to do an averaging
of the flow variables over the $z$ direction. We define
\begin{equation}
\label{e14}
\big{\langle} \psi (x,z) \big{\rangle} = \frac{1}{h(x)}
\int^{h(x)}_{0} \psi (x,z)\, \mathrm{d}z
\end{equation}
where the averaged quantity in the angled brackets becomes a function
of $x$ only. Under the assumption of a flat free surface such that
$w(x,z)=0$ at $z=h(x)$, and along with the use of Eq.(\ref{e11}), 
we can readily show that
\begin{equation}
\label{e15}
\bigg{\langle} w\frac{\prt u}{\prt z} \bigg{\rangle} = \bigg{\langle}
u\frac{\prt u}{\prt x} \bigg{\rangle}
\end{equation}
and
\begin{equation}
\label{e16}
\bigg{\langle}\frac{\prt^2u}{\prt z^2}\bigg{\rangle}
= - \frac{1}{h(x)} \frac{\prt u}{\prt z}\bigg{\vert}_{z=0}
\end{equation}
We now make the approximations,
\begin{equation}
\label{e17}
\bigg{\langle} u\frac{\prt u}{\prt x} \bigg{\rangle} = \alpha \langle
u \rangle \frac{\prt \langle u \rangle}{\prt x}
\end{equation}
and
\begin{equation}
\label{e18}
\frac{\prt u}{\prt z}\bigg{\vert}_{z=0} = \beta \frac{\langle u
\rangle}{h(x)}
\end{equation}
where $\alpha$ and $\beta$ are numbers of $\mathcal{O}(1)$ and 
they depend upon the velocity profile. For a parabolic profile 
$\alpha = 6/5$ and $\beta = 3$. It is easy to check that the 
parameters $\alpha$ and  $\beta$ are strictly constants when we make, 
following Watson \cite{wat64}, the reasonable scaling ansatz that
$u(x,z)=U(x)F[z/h(x)]$. The fact that our experiment will determine a
combination of $\alpha$ and $\beta$ is important regarding the nature
of the vertical profile of the velocity field over which we average.

With the above approximations and identity, and writing 
$\langle u \rangle$ as $v$, Eq.(\ref{e12}) becomes
\begin{equation}
\label{e19}
2 \alpha v \frac{\mathrm{d}v}{\mathrm{d}x} = 
-g \frac{\mathrm{d}h}{\mathrm{d}x} -\beta\nu\frac{v}{h^2}
\end{equation}
From Eq.(\ref{e13}) we get $Lvh=Q$, which we can use to eliminate $v$
from Eq.(\ref{e19}) and get
\begin{equation}
\label{e20}
\bigg[2 \alpha \bigg(\frac{Q}{L} \bigg)^2 -gh^3 \bigg]
\frac{\mathrm{d}h}{\mathrm{d}x} = \beta\nu\frac{Q}{L}
\end{equation}
The derivation of the above equation may look somewhat heuristic, 
but it has been much more systematic to the extent that it avoids
the inclusion of an artificial friction loss. The structure of 
this equation, however, is quite similar to what is well known in 
hydraulic jump literature \cite{gra87}, to derive which, the conventional 
recourse has been to use the Bernoulli equation supplemented by a
friction loss. In our derivation of Eq.(\ref{e20}), the approximations 
made on the Navier-Stokes equation have been clearly delineated, and the 
resulting profile has been seen to be sensitive to the velocity field in 
a fashion that can be experimentally probed. This equation shows that 
if $\nu=0$, then $h(x)$ would be a constant. Therefore, without
viscosity the hydraulic jump can be obtained only if we explicitly
seek a solution with the jump extraneously imposed, which is the way
it has been conventionally treated in text books. The jump occurs
when $\mathrm{d}h/\mathrm{d}x$ displays singular behaviour at
$h=h_{\mathrm{j}}$ such that 
$gh_{\mathrm{j}}^3 = 2 \alpha (Q/L)^2$. 
The advantage  of Eq.(\ref{e20}) is that it can be exactly
integrated. This gives,
\begin{equation}
\label{e21}
gh \left(h_{\mathrm{j}}^3 - \frac{h^3}{4} \right)   =
\beta\nu\frac{Q}{L}x + \mathcal{C}
\end{equation}
where $\mathcal{C}$ is a constant of integration. The position
$x_{\mathrm{j}}$ of the jump is obtained by setting $h=h_{\mathrm{j}}$
in Eq.(\ref{e21}). This yields
\begin{equation}
\label{e22}
x_{\mathrm{j}} =\frac{3}{\beta}\bigg( \frac{\alpha ^4}{4} \bigg)^{1/3}
\bigg( \frac{Q}{L} \bigg)^{5/3} \nu^{-1} g^{-1/3} + \tilde{\mathcal{C}}
\end{equation}
where $\tilde{\mathcal{C}}$ is some other constant.

The complete profile of $h(x)$ is described by Eq.(\ref{e21}), but it
should be noted that it gives multiple $h(x)$ for a given $x$. However, 
for values of $h(x)$ which are even moderately less that $h_{\mathrm{j}}$, 
the profile according to Eq.(\ref{e21}) can be approximated as
\begin{equation}
\label{e23}
h(x) \sim \left( \frac{\beta\nu}{2\alpha}\frac{L}{Q} \right) x
\end{equation}
which tells us that for small $x$ the height of the liquid layer
increases linearly. This feature is markedly different from the case
of the radial spread of a liquid stream on a horizontal plate in
which the height gradually decreases in the region within the 
jump \cite{ot66}.

With the help of Eq.(\ref{e21}) we are now in a position to get a 
physical picture of the jump. We introduce the dimensionless variables
$H=h/h_{\mathrm{j}}$ and $X=x/\tilde{x}_{\mathrm{j}}$ in which, 
$\tilde{x}_{\mathrm{j}} = x_{\mathrm{j}} - \tilde{\mathcal{C}}$.
We thus obtain a dimensionless form of Eq.(\ref{e21}) as
\begin{equation}
\label{e24}
4H-H^4 = 3 \left(X - D \right)
\end{equation}
with $D$ being a dimensionless constant. Differentiating Eq.(\ref{e24}) 
will readily show that at $H=1$, the function $H(X)$ will have a vertical 
tangent --- something that can be conceived of as a discontinuity in the
physical flow. Our analysis is restricted to the range $X \geq 0$. The 
point along the $X$-axis, where the flow will encounter this discontinuity, 
will be determined by the inner boundary condition. To show this, we 
set the condition $H=0$ at an arbitrary $X=X_\mathrm{in}$, to get 
$D=X_\mathrm{in}$. When $X=1+X_\mathrm{in}$, we will have $H=1$, which
gives us a clear indication that the value of $X_\mathrm{in}$ will
determine the position of the discontinuity. 

\begin{figure}
\begin{center}
\resizebox{1.0 \hsize}{!}{\includegraphics{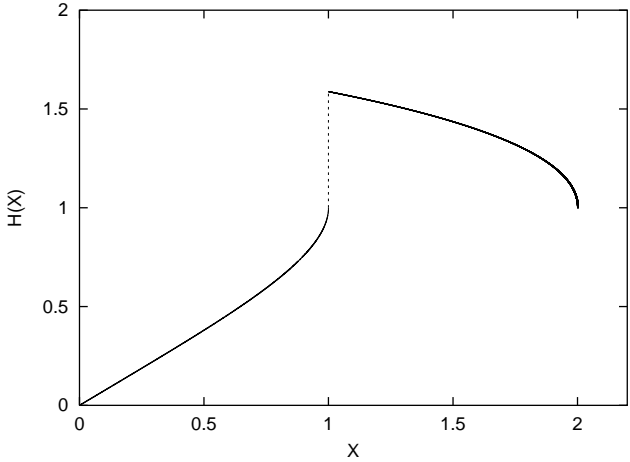}}
\end{center}
\caption{Fitting the inner and outer solutions via a shock.
The inner solution corresponds to $X < 1$, while the outer
solution corresponds to $X > 1$. Each solution has been
determined by a different boundary condition. The vertical
dotted line (the shock) joins the two solutions.}
\label{Fig4}
\end{figure}

A generic inner boundary condition is that $H=0$ at $X=0$ (ignoring a 
small initial non-zero value of $H$ at $X=0$), giving us 
$X_\mathrm{in}=D=0$. The solution corresponding to this particular
boundary condition is represented by the lower branch in 
Fig.\ref{Fig4}. In the dimensionless notation that we have introduced, 
this means a profile proceeding from $X=0$ to $X=1$ and rising from 
$H=0$ to $H=1$. Since the inner solution diverges infinitely at $X=1$, 
the flow has to physically go beyond this point by the fitting of 
the lower branch to the upper branch in Fig.\ref{Fig4}, through the 
discontinuity. We treat this discontinuity as the jump in the flow. 

The integration constant $D$ for the upper branch is to be fixed by 
the outer boundary condition. This is determined by the physical 
requirement that $H=1$ (infinite slope) at $X=X_{\mathrm{e}}$, with 
$X_{\mathrm{e}}$ corresponding to a position slightly left to the 
edge of the channel where the liquid falls off. We cannot extend the 
solution to the edge due to singular flow at the outlet. For the 
outer boundary condition stated above we will get the solution 
$4H-H^4=3 \left(X-X_{\mathrm{e}} +1 \right)$. At the position of 
the jump, i.e. at $X=1$, the outer solution will give the magnitude
of the jump, $H_\mathrm{J}$, making it evident
that $H_\mathrm{J}$ has a dependence on $X_{\mathrm{e}}$. It 
should be a worthwhile exercise to study this dependence. 

The roots of $H_\mathrm{J}$ can be determined by solving the 
biquadratic equation given by 
$H_\mathrm{J}^4 -4H_\mathrm{J}+3 \left(2-X_{\mathrm{e}} \right) =0$. 
To that end, we will first have to carry out a transformation with 
the help of a new variable $\eta$, to write 
$\left( H_\mathrm{J}^2 + \eta \right)^2 = 2 \eta H_\mathrm{J}^2 
+ \eta^2 + 4H_\mathrm{J} - 3 \left(2-X_{\mathrm{e}}\right)$, and 
then require that by a suitable choice of $\eta$, the right hand
side will be rendered a perfect square. This requirement will mean
that the discriminant of the quadratic in $H_\mathrm{J}$ on the right 
hand side should vanish, to ultimately yield the auxiliary cubic 
equation $\eta^3 - 3\eta \left(2-X_{\mathrm{e}}\right) -2 = 0$. 

The discriminant, $\Delta$, of this cubic equation will be given by
$\Delta = 1 - \left(2-X_{\mathrm{e}}\right)^3$, and it is quite easy to 
see that for $X_\mathrm{e} > 1$, there will always be a positive value
for $\Delta$. Therefore, $\eta$ can have only one real root, $\eta_0$, 
given by $\eta_0 = \left(1 + \sqrt{\Delta} \right)^{1/3} + 
\left(1 - \sqrt{\Delta} \right)^{1/3}$. To solve for $H_\mathrm{J}$,
we now use this value of $\eta_0$ in the biquadratic equation 
$\left( H_\mathrm{J}^2 + \eta_0 \right) = \pm \sqrt{2 \eta_0} 
\left(H_\mathrm{J} + \eta_0^{-1}\right)$. Of course, there will be 
four roots of $H_\mathrm{J}$, but the real roots should correspond
to the choice of the positive sign. Solving the relevant quadratic
equation and choosing to keep only the physically meaningful positive
root, will give us the solution of $H_\mathrm{J}$ as 
\begin{equation}
\label{e25}
H_\mathrm{J} = \sqrt{\frac{\eta_0}{2}} + 
\sqrt{\left(\frac{2}{\eta_0}\right)^{1/2} -\frac{\eta_0}{2}}
\end{equation}
An interesting conclusion that can be derived from the relation of
the jump height above is that for $X_\mathrm{e} \gg 1$, the maximum
height of the jump will asymptotically be given by
$H_\mathrm{J} \simeq \left(3 X_\mathrm{e} \right)^{1/4}$. 

As a specific case we set $X_\mathrm{e}=2$, and then determining
the values of both $\Delta$ and $\eta_0$, we see that 
$H_\mathrm{J} = 4^{1/3}$. This result could alternatively be
derived directly from Eq.(\ref{e24}) by the use of the boundary
condition, $H=1$, at $X=2$. This will give $D=1$, and for this
particular boundary condition, the resulting outer solution has
been plotted in Fig.\ref{Fig4}. In this instance the fractional 
change in height at the jump position would
be $4^{1/3} - 1 \simeq 0.59$. 

The fitting of the upper and the lower branches (each determined by a 
different boundary condition) has to be done via a shock, by requiring 
that the mass and momentum fluxes remain unchanged through the 
discontinuity --- a condition that may be derived from the inviscid 
equations. From Eqs.(\ref{e1}) and (\ref{e2}), with both $h_1$ and $h_2$ 
scaled by $h_{\mathrm{j}}$ to give $H_1$ and $H_2$, we get what 
Bohr et al. \cite{bdp93} call the jumpline
$H_1 H_2 \left( H_1 + H_2 \right) = \alpha^{-1}$. 
The jump takes place when $H$ of the viscous solution in the lower
branch of Fig.\ref{Fig4} becomes equal to $H_1$ of the inviscid 
jumpline for a given value of $\alpha$. In general, the shock fitting 
leads to a jump position actually slightly left of $X=1$.

\section{Time-dependence in the Channel Flow}
\label{sec4}

So far our analysis has been carried out in terms of the steady flow 
equations only. We now consider explicit time-dependence in both 
the equation 
of continuity and the Navier-Stokes equation. This will necessitate 
the time-dependent generalisation of both the flow variables $h$, the
local height of the flow, and $v$, the vertically integrated average
flow velocity. The resultant dynamic equations may then be written as
\begin{equation}
\label{e26}
\frac{\prt h}{\prt t} + \frac{\prt}{\prt x}\left(vh \right) = 0
\end{equation}
and
\begin{equation}
\label{e27}
\frac{\prt v}{\prt t} + v\frac{\prt v}{\prt x} + g\frac{\prt h} {\prt
x} = - \nu \frac{v}{h^2}
\end{equation}
respectively.

For the purpose of carrying out a linear stability analysis of the 
flow in real time, it will be convenient
for us to work with a new variable which we define as $f=vh$. The new
variable $f$ can be physically associated with the time-dependent
volumetric flow rate, and its steady solution, as can be seen from
Eq.(\ref{e26}), is a constant.
We use solutions of the form $v(x,t) = v_0(x) + v^{\prm}(x,t)$ and
$h(x,t) = h_0(x) + h^{\prm}(x,t)$, in which the subscript $0$
indicates steady solutions, while the primed quantities are
time-dependent perturbations about those steady solutions. Linearising
in these fluctuating quantities gives us the fluctuation of $f$ about
its constant steady value $f_0 = v_0 h_0 = Q/L$, as
\begin{equation}
\label{e28}
f^{\prm}=v_0 h^{\prm} + h_0 v^{\prm}
\end{equation}

In terms of $f^{\prm}$, we can then write from Eq.(\ref{e26}),
\begin{equation}
\label{e29}
\frac{\prt h^{\prm}}{\prt t} = - \frac{\prt f^{\prm}}{\prt x}
\end{equation}
and this in turn, along with Eq.(\ref{e28}), gives,
\begin{equation}
\label{e30}
\frac{\prt v^{\prm}}{\prt t} = \frac{1}{h_0}
\left(\frac{\prt f^{\prm}}{\prt t} \right) + 
\frac{v_0}{h_0}\left( \frac{\prt f^{\prm}}{\prt x}\right) 
\end{equation}
A further partial derivative of Eq.(\ref{e30}) with respect to time 
yields
\begin{equation}
\label{e31}
\frac{\prt^2 v^{\prm}}{\prt t^2} = \frac{\prt}{\prt t}
\left[\frac{1}{h_0} \left(\frac{\prt f^{\prm}}{\prt t} \right)\right] 
+ \frac{\prt}{\prt t}\left[ 
\frac{v_0}{h_0}\left( \frac{\prt f^{\prm}}{\prt x}\right)\right]
\end{equation}
The significance of the form in which we have kept Eq.(\ref{e31}), 
will be apparent soon. 
Linearising in the perturbed quantities in Eq.(\ref{e27}) gives
\begin{equation}
\label{e32}
\frac{\prt v^{\prm}}{\prt t} + \frac{\prt}{\prt x}\left(v_0
v^{\prm}\right) + g \frac{\prt h^{\prm}}{\prt x} = -
\frac{\nu}{h_0^2}\bigg( v^{\prm} -2 v_0 \frac{h^{\prm}}{h_0} \bigg)
\end{equation}
which in turn, upon partially differentiating with respect to $t$, 
yields
\begin{eqnarray}
\label{e33}
\frac{\prt^2 v^{\prm}}{\prt t^2} + \frac{\prt}{\prt x}\left(v_0
\frac{\prt v^{\prm}}{\prt t}\right) &+& g \frac{\prt}{\prt x}
\left(\frac{\prt h^{\prm}}{\prt t}\right) \nonumber \\
&=& - \frac{\nu}{h_0^2}
\left[ \frac{\prt v^{\prm}}{\prt t} -2 \frac{v_0}{h_0} 
\left(\frac{\prt h^{\prm}}{\prt t}\right) \right] 
\end{eqnarray}
In Eq.(\ref{e33}) above, we substitute for the first and the second
order time derivatives of $h^{\prm}$ and $v^{\prm}$ from Eqs.(\ref{e29}) 
to (\ref{e31}). This will lead to the result
\begin{eqnarray}
\label{e34}
& & \frac{\prt}{\prt t} \left[\frac{1}{h_0} 
\left( \frac{\prt f^{\prm}}{\prt t}\right)\right] 
+ \frac{\prt}{\prt t} \left[\frac{v_0}{h_0}
\left( \frac{\prt f^{\prm}}{\prt x}\right)\right]
+ \frac{\prt}{\prt x} \left[\frac{v_0}{h_0}
\left( \frac{\prt f^{\prm}}{\prt t}\right)\right] \nonumber \\
&+&  \frac{\prt}{\prt x} \left[\frac{1}{h_0}
\left(v_0^2 - gh_0 \right) \frac{\prt f^{\prm}}{\prt x}\right]
= - \frac{\nu}{h_0^3}\left(\frac{\prt
f^{\prm}}{\prt t} + 3 v_0 \frac{\prt f^{\prm}}{\prt x} \right)
\end{eqnarray}
At this juncture it should be most instructive 
for us to examine Eq.(\ref{e34})
in its inviscid limit, i.e. when $\nu=0$. In connection with this, it 
should also be very much worth our while to consider some recent 
studies \cite{su02,blv05,vol05} which have revealed a close analogy
between the propagation of a wave in a moving fluid and of light in
curved space-time. In particular Sch\"{u}tzhold and Unruh have shown 
how gravity waves in a shallow layer of liquid are governed by the 
same wave equation as for a scalar field in curved space-time \cite{su02}. 
For an inviscid, incompressible and irrotational flow, these authors 
have prescribed the flow velocity to be the gradient of a scalar potential.
Perturbing this potential about its background value, under the 
restricted condition of the flow height being a constant, leads to
an effective metric of the flow, in which the velocity of gravity 
waves replace the speed of sound in sonic analogs that faithfully
reflect features seen in general relativistic studies \cite{su02,blv05}.

We now proceed to demonstrate that our perturbative analysis of what 
is essentially a dissipative system (since it includes viscosity), 
will, in its inviscid limit, deliver the same metric obtained by 
Sch\"{u}tzhold and Unruh from their purely inviscid model. It 
must be stressed here that our choice of perturbing the flow rate 
is paticularly expedient, since conservation of matter holds good 
even in a system that undergoes viscous dissipation. It is to be 
further noted that the background velocity and flow height in 
our treatment are in general functions of space and not constants. 
Having made these observations to emphasise the greater generality
and exactitude of our approach, we can then extract the inviscid
terms from Eq.(\ref{e34}) by setting $\nu=0$, to ultimately render 
these terms into a compact formulation that looks like \cite{blv05}  
\begin{equation}
\label{e35}
\prt_\mu \left( {\mathrm{f}}^{\mu \nu} \prt_\nu 
f^{\prm}\right) = 0
\end{equation}
in which, we make the Greek indices run from $0$ to $1$, with the
identification that $0$ stands for $t$ and $1$ stands for $x$. 
An inspection of the terms in the left hand side of Eq.(\ref{e34})
--- all of them divided by the constant $g$ --- will then enable us
to construct the symmetric matrix
\begin{equation}
\label{e36}
{\mathrm{f}}^{\mu \nu } = \frac{1}{gh_0}
\pmatrix { 1 & v_0 \cr
v_0 & v_0^2 - gh_0}
\end{equation}
Now in terms of the metric ${\mathrm{g}}_{\mu \nu}$, the d'Alembertian 
for a scalar in curved space, is given by \cite{blv05}
\begin{equation}
\label{e37}
\triangle \psi \equiv \frac{1}{\sqrt{-\mathrm{g}}}
\prt_\mu \left({\sqrt{-\mathrm{g}}}\, {\mathrm{g}}^{\mu \nu} \prt_\nu
\psi \right) 
\end{equation} 
in which $\mathrm{g}^{\mu \nu}$ is the inverse of the matrix 
implied by ${\mathrm{g}}_{\mu \nu}$. 
Under the equivalence that ${\mathrm{f}}^{\mu \nu } =
\sqrt{-\mathrm{g}}\, {\mathrm{g}}^{\mu \nu}$, and therefore, 
$\mathrm{g} = \det \left({\mathrm{f}}^{\mu \nu }\right)$, we can 
immediately set down an effective metric 
\begin{equation}
\label{e38}
\mathrm{g}^{\mu \nu}_{\mathrm{eff}}
= \pmatrix { 1 & v_0 \cr
v_0 & v_0^2 - gh_0}
\end{equation}
that is entirely identical to the one obtained by Sch\"{u}tzhold and 
Unruh, following whom, the inverse effective metric, 
$\mathrm{g}_{\mu \nu}^{\mathrm{eff}}$,
can also be easily obtained from Eq.(\ref{e38}). This shall identify 
$v_0^2 = gh_0$ as the ergosphere condition on the horizon of either a 
black hole or a white hole, depending on the direction of the flow. In 
the case of the two-dimensional circular hydraulic jump, as Volovik has
pointed out \cite{vol05}, the jump condition can be closely related
to the horizon of a white hole, a surface that nothing can penetrate. 
This analogy is entirely apt for the channel flow that we are studying 
here, considering the direction in which the flow proceeds.  

Introduction of viscosity in the flow does affect the idealised inviscid
conditions in the vicinity of the white hole horizon. Sch\"{u}tzhold 
and Unruh themselves have treated viscosity as a small adjunct effect 
on the inviscid flow, and have concluded that although viscosity will 
introduce a thin boundary layer, the flow outside it shall very well 
be governed by inviscid conditions, and that the basic properties of 
gravity waves will not be drastically affected. In our presentation
of the perturbative analysis, we have systematically included viscosity 
in our governing equations, which has finally led to Eq.(\ref{e34}). 
It is quite evident that the presence of viscosity disrupts the 
precise symmetry of the inviscid conditions described Eq.(\ref{e35}). 
This obviously implies that the clear-cut horizon 
condition that one obtains from the inviscid limit, will be affected.
However, this effect, for small viscosity, as Sch\"{u}tzhold and Unruh 
have argued, cannot be too drastic. This is very much in conformity 
with our observation in the previous section that fitting the inner 
and outer solutions through a shock will shift the position of the jump
only slightly. One way or the other, the most important feature to emerge 
from the analogy of white hole horizon shall remain qualitatively 
unchanged, namely, that a disturbance propagating upstream from the 
subcritical
flow region (where $v_0^2 < gh_0$) cannot penetrate through the horizon 
(where $v_0^2 = gh_0$) into the supercritical region of the flow (where
$v_0^2 > gh_0$). As we shall see shortly, this property of the flow
will have a very crucial bearing on a physical picture that we shall
construct to explain the formation of the hydraulic jump. 

For our purposes it should also be important to study the 
behaviour of the perturbation $f^{\prm}$. To that end we go back 
to Eq.(\ref{e34}) and recast it in a slightly altered form that 
looks like
\begin{eqnarray}
\label{e39}
\frac{{\prt}^2 f^{\prm}}{\prt t^2} + 2 \frac{\prt}{\prt x}\left(v_0
\frac{\prt f^{\prm}}{\prt t}\right) &+& \frac{1}{v_0}\frac{\prt}{\prt
x} \left[v_0 \left(v_0^2 - gh_0 \right) \frac{\prt f^{\prm}}{\prt
x}\right] \nonumber \\ &=& - \frac{\nu}{h_0^2}\left(\frac{\prt
f^{\prm}}{\prt t} + 3 v_0 \frac{\prt f^{\prm}}{\prt x} \right)
\end{eqnarray}
Using a solution of the type
$f^{\prm}(x,t) = p(x)\exp(- \mathrm{i} \omega t)$ in Eq.(\ref{e39})
gives an expression that is to be further multiplied by $v_0 p$ 
throughout, to finally deliver a quadratic equation in $\omega$ 
that is of the form
\begin{eqnarray}
\label{e40}
&-& \omega^2 \left(v_0 p^2 \right) - \mathrm{i} \omega 
\left[ \frac{{\mathrm{d}}}{{\mathrm{d}x}} \left(v_0 p \right)^2 
+ \nu \frac{v_0 p^2}{h_0^2} \right] \nonumber \\
&+& p \frac{\mathrm{d}}{\mathrm{d}x} \left[v_0 \left(v_0^2 - gh_0 
\right) \frac{\mathrm{d}p}{\mathrm{d}x} \right]
+ 3 \nu \frac{v_0^2 p}{h_0^2} 
\left(\frac{\mathrm{d}p}{\mathrm{d}x}\right)=0 
\end{eqnarray}
To have any idea of how the perturbation behaves in time, we treat
it as a standing wave. For that purpose we will have to integrate 
the above equation between two chosen boundaries, at which the 
perturbation will be constrained to vanish at all times. Between 
these two boundaries the flow should be continuous. Since we are 
aware that the jump itself is a discontinuity in the flow, we will have 
to choose the boundaries to be on one side of the jump only, although 
Eq.(\ref{e40}) itself holds true over the entire range of the flow.
We have already acquainted ourselves with the fact that a perturbation
in the subcritical region will remain confined to this region only. 
Besides, in this region the flow would have entirely lost its laminar 
character, and therefore, would be most suited for us to derive some 
physical insight about the behaviour of the perturbation and the 
influence of viscosity on that. Therefore, we confine our analysis 
to the subcritical region of the flow only. 
As for the boundaries of the perturbation, one of them can be the 
outer boundary of the steady flow itself, while the inner boundary
may be chosen to be very close to the jump. In this regard we treat
the jump itself as a boundary wall where all velocity 
and height fluctuations decay out completely. Under these conditions
an integration of Eq.(\ref{e40}) leads to 
\begin{eqnarray}
\label{e41}
{\omega}^2 \int v_0 p^2 \, \mathrm{d}x &+& \mathrm{i} \omega \nu \int
\frac{v_0 p^2}{h_0^2} \, \mathrm{d}x - 3 \nu \int \left(
\frac{v_0}{h_0} \right )^2 p \frac{\mathrm{d}p}{\mathrm{d}x} \,
\mathrm{d}x \nonumber \\ &+& \int v_0 \left (v_0^2 - gh_0 \right )
\left (\frac{\mathrm{d}p}{\mathrm{d}x} \right )^2 \,  \mathrm{d}x = 0
\end{eqnarray}
which is a result that has been arrived at by carrying out the 
integration by parts, and then by imposing the requirement that 
all the integrated ``surface" terms vanish at the two boundaries.

Under inviscid conditions, $\omega$ will have a purely real solution,
and therefore the perturbation will be a standing wave with a constant
amplitude in time. However, the dissipative presence of viscosity
will result in the perturbation being damped out, and will restore 
the system towards a stable configuration. The consequent
time-decay of the amplitude of the perturbation will be of the form
$\exp (- \nu t/2 h_0^2)$. This also gives a time scale for viscous 
dissipation, which, to an order-of-magnitude, is given by 
$t_{\mathrm{visc}} \sim h_0^2 / \nu$. 

It is now important to appreciate that viscous drag in the fluid will 
also dissipatively slow down the flow on this very same time scale
$t_{\mathrm{visc}}$. The information of an advanced layer of fluid
slowing down has to propagate upstream to preserve the smooth
continuity of the fluid flow. This propagation, however, cannot 
happen any faster than the speed of the surface gravity waves,
$(gh_0)^{1/2}$, and in the region where $v_0 > (gh_0)^{1/2}$, 
i.e. in the supercritical region, no
information therefore can propagate upstream \cite{bpw97}. This is
also what we should entirely expect from the modelling of the jump 
as the horizon of an impenetrable white hole. 
So a stream of fluid that has arrived later, after having passed
through the supercritical region, moves on ahead, yet unhindered 
and uninformed, till its speed becomes comparable with the speed of 
the surface gravity waves, and only then does any knowledge about 
a ``barrier" ahead catches up with the fluid. We now define a dynamic 
time scale, $t_{\mathrm{dyn}} \sim {x}/{v_0}$, on which the bulk flow 
proceeds. If we set $t_{\mathrm{visc}} \simeq t_{\mathrm{dyn}}$ 
with the additional requirement $v_0 \simeq (gh_0)^{1/2}$, we get the 
condition for the ``news" of the viscous slowing down finally catching 
up with the bulk flow itself. With the additional constraint that we have 
from the continuity equation, i.e. $v_0 h_0 = Q/L$, we can then derive 
a scaling relation for length which is entirely identical to the 
scaling dependence of the position of the hydraulic jump,
\begin{equation}
\label{e42}
x_{\mathrm{j}} \sim \left(\frac{Q}{L} \right)^{5/3} {\nu}^{-1} g^{-1/3}
\end{equation}
as given by Eq.(\ref{e22}) --- a result that we have already obtained
from our study of the stationary flow equations.

The crux of the argument that emerges from our analysis is that 
for the formation of the hydraulic jump, the two time scales,
$t_{\mathrm{visc}}$ and $t_{\mathrm{dyn}}$, would have to match each
other closely when the Froude number, $\mathcal{F}$, is close to
unity. Under these conditions, a layer of fluid arriving lately is 
confronted with a barrier formed by a layer of fluid moving ahead with 
an abrupt slowness. This slowly moving layer of fluid flowed past earlier 
in time, and at far distances it has been retarded considerably by 
viscous drag. Since in this situation there could not be an indefinite 
accumulation of the fluid,
and since continuity of the fluid flow has to be preserved, the newly
arrived fluid layer slides over the earlier viscosity dragged slowly
moving layer of the fluid, and what we see is a sudden increase in 
the height of the fluid layer --- a hydraulic jump. This gives a
conceivable physical basis for understanding how crucial a factor
viscosity is behind the formation of the jump, and in connection  
with this physical picture, it is also worthwhile to ponder the 
possibility that the viscosity-induced boundary layer of the flow 
gradually increases in thickness and the jump occurs at that distance, 
where the boundary layer pervades the entire height of the thin layer 
of the flowing fluid, i.e. from $z=0$ to $z=h(x)$.

While dwelling on the issue of the physical picture of the jump
formation, we are tempted to adduce a related astrophysical 
analogy : the formation of hot spots in galaxies. Gaseous jets
emanating from galaxies encounter resistence from the intergalactic
medium. As a result the tip of the jet is slowed down in comparison
with its bulk. This causes energy to accumulate at the tip, and this
is a likely explanation for the formation of a hot spot. The gas
flow in the jet proceeds at supersonic speeds, but on coming close
to the hot spot, the flow experiences an abrupt deceleration. The
information of this sudden braking is not conveyed upstream, since
no sound wave can move against the supersonic flow. This causes a 
shock wave to form across the jet \cite{begrees}. In this particular
astrophysical picture, one might discern much similarity of principle
with our physical arguments behind the formation of the hydraulic jump
in the channel flow.

Another significant point of which we should like to make mention is
that instead of Eq.(\ref{e27}) we could have chosen to use the dynamic
generalisation of Eq.(\ref{e19}) with the dimensionless constants
$\alpha$ and $\beta$ included, but to derive the particular form of 
the perturbation equation as given by Eq.(\ref{e34}), we would have 
to set $\alpha = 1/2$. This argument possibly has an important bearing
on the issue of the velocity profile of the flow.

\section{Experimental Results}
\label{sec5}

A relatively simple experiment, using water, was carried out to 
substantiate our theoretical propositions. Our objectives were 
two-fold. First, to verify the linear growth of the surface height
of the flow at small distances, and secondly, to verify the scaling
of the jump position as given by Eq.(\ref{e42}). It has been 
satisfying to note that our theory --- presented heuristically in
parts --- has very much been borne out within the limited objectives
of our experiment. 

The experimental apparatus has already been illustrated schematically
in Fig.\ref{Fig1}. The water flows in a streamline motion through 
an open rectangular channel of width $L$. An arrangement has been 
made such that the water will flow down an inclined channel, with 
an inclination of $60^{\circ}$, and then from the edge $AB$, will 
start its one-dimensional motion in the horizontal channel whose 
length is $70 \, \mathrm{cm}$ and width is $9.1 \, \mathrm{cm}$. A 
rectangular box is attached to the base of the inclined channel. Water 
comes from a tap to the rectangular box. There is a slit in the box 
through which water flows down the channel. The purpose of the inclined 
channel is to make the flow laminar from the beginning of the horizontal 
channel. Although this will introduce a small horizontal component to
the flow, in so far as we would be interested in the scaling relations
only, any small extraneous addition to the flow should not too 
drastically affect our observations. In any case, preserving the 
laminarity of the flow should be necessary for the unambiguous 
recording of data. For that purpose, the height of the water layer has 
been measured with a travelling microscope. The flow has been viewed 
through the transparent perspec wall of the horizontal channel. At each 
value of $x$, we noted down the vertical positions of the bottom of the 
flow in contact with the bed of the channel, and of the free surface of 
the flowing water (both of which were easily identifiable in the field 
of view of the microscope). The difference of the two readings gave $h(x)$. 

Viewed from the top of the channel, the jump itself has been seen to 
present a curved front across the width (i.e. along the $y$ axis) of 
the channel. This is because the flow in contact with the boundary 
walls of the channel (at $y=0$ and $y=L$ respectively) has been dragged 
down by viscosity. The position of the jump, $x_{\mathrm{j}}$, is 
actually an average value measured over this lateral curved profile. 

To estimate the volumetric flow rate, we adopted the simple recourse of 
collecting the water falling off at the outer edge of the channel, and 
then of measuring the volume of the water collected for various intervals 
of time. The average of all these readings has been taken to determine $Q$. 
The steadiness of the flow has also been confirmed by this approach. Values 
of $X_{\mathrm{e}}$, at the outer edge of the channel (discussed in 
Section \ref{sec3}), range between $1.5$ and $2.8$ in our experimental 
set-up.

\begin{figure}
\begin{center}
\resizebox{1.0 \hsize}{!}{\includegraphics{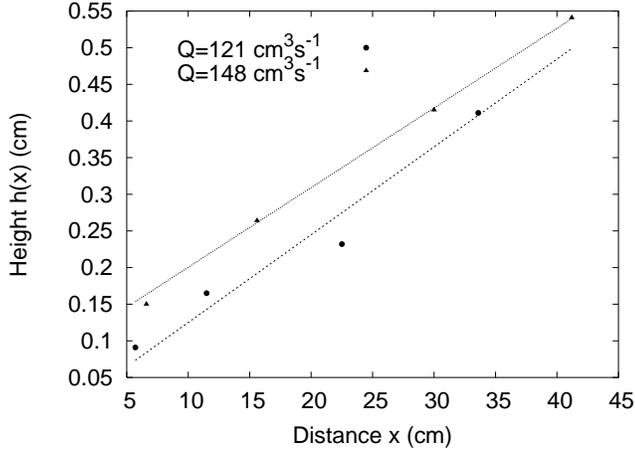}}
\end{center}
\caption{The height profile $h(x)$ before the jump for different
values of $Q$, the volumetric flow rate.}
\label{Fig5}
\end{figure}

\begin{figure}
\begin{center}
\resizebox{1.0 \hsize}{!}{\includegraphics{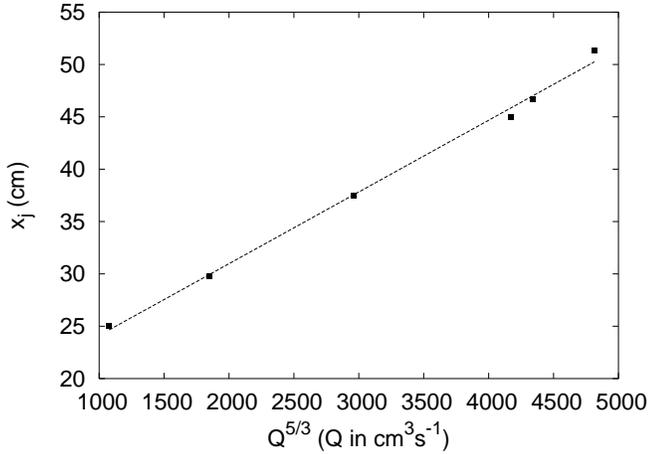}}
\end{center}
\caption{The position $x_{\mathrm{j}}$ of the jump for different
volumetric flow rates $Q$.}
\label{Fig6}
\end{figure}

We observed a very slow rise in $h(x)$ for a while and
then a major jump. That the rising profile of $h(x)$ will be linear, 
as we might rightly expect on the basis of Eq.(\ref{e23}), has been 
shown in Fig.\ref{Fig5} for two different values of $Q$. We also show 
the scaling dependence of the jump position $x_{\mathrm{j}}$ on 
$Q^{5/3}$ in Fig.\ref{Fig6}. The predicted linearity from Eq.(\ref{e42}) 
has been depicted quite clearly here. It is obvious that once the $5/3$
power dependence of $x_{\mathrm{j}}$ on $Q$ has been established 
experimentally, the dependence on $\nu$ and $g$, as Eq.(\ref{e42}) shows, 
must follow even on the basis of elementary dimensional considerations. 
So our theoretical scaling law stands vindicated by our 
simple experiment. 

\begin{figure}
\begin{center}
\resizebox{1.0 \hsize}{!}{\includegraphics{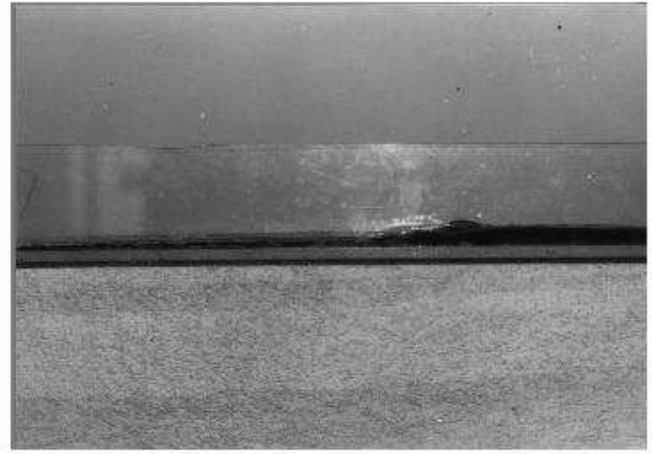}}
\end{center}
\caption{A sideview of jump region with the flow proceeding from the
left to the right. The flow appears black in the photograph, because 
it has been coloured with a red dye. The jump is clearly discernable, 
as also are the growth and decay profiles of the surface height, before 
and after the jump, respectively. }
\label{Fig7}
\end{figure}

\begin{figure}
\begin{center}
\resizebox{1.0 \hsize}{!}{\includegraphics{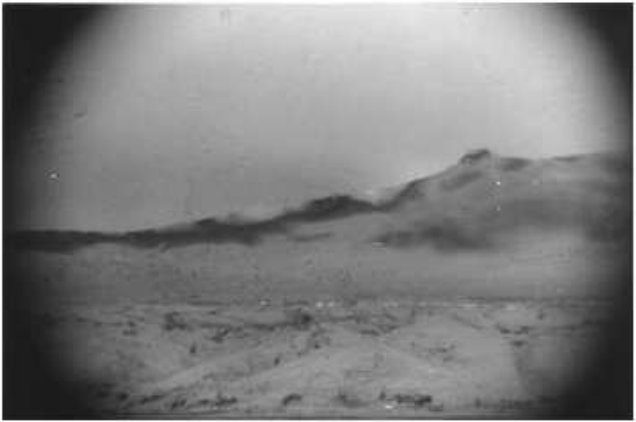}}
\end{center}
\caption{A close-up sideview of the jump region.}
\label{Fig8}
\end{figure}

More to this point, we also furnish two photographs of the cross-section
of the flow with the jump included. We show a long distance snapshot of 
the flow in Fig.\ref{Fig7}. The flow has been dyed red to make it more 
prominent, when viewed through the transparent perspec wall. The jump is 
very much discernable in this photograph, but we must also draw attention 
to the slow linearised growth of the height of the flow much before the 
jump, followed by its much more rapid growth immediately in front of the 
jump. This is further followed by a small decrease in the flow level in 
the region beyond the jump. Qualitatively this is what we should expect 
on the basis of the plot in Fig.\ref{Fig4}. That the variation of the flow 
height we see in the photograph, is not as pronounced as Fig.\ref{Fig4} 
would impress upon us, is because of the fact that the axes in 
Fig.\ref{Fig4} have been scaled in terms of the flow constants, while 
the variation of the flow height in the actual photograph proceeds on 
the scale of CGS units. The jump appears much more prominently in the 
close-up view of the jump region, shown in Fig.\ref{Fig8}. 

From Table \ref{tab1} it is seen that the jump remains almost
constant within an experimental error of about $6\%$. The combined
effect of the theoretical analysis and the experimental results leads
us to believe that the magnitude of the jump in a shallow layer flow
will show no drastic variation. 

The $z$ profile of the velocity field $u(x,z)$ also 
calls for some comments.
The slope in Eq.(\ref{e22}) depends on the numbers $\alpha$ and
$\beta$ according to the combination $\alpha^{4/3}/\beta$, which, for
a parabolic profile ($\alpha=6/5$ and $\beta=3$) is about $0.425$. 
This ultimately gives a theoretical estimate of the slope to be 
approximately $0.204$. But the experimental data as plotted
in Fig.\ref{Fig6} show that the actual slope (nearly $0.007$) is much
smaller. This leads us to infer that the profile is much steeper than
parabolic near the plate and much flatter near the free surface.

\begin{table}
\begin{center}
\begin{tabular}{|p{0.6in}|p{0.6in}|p{0.6in}|p{0.6in}|}  \hline\hline
$Q$ & $h_1$ & $h_2$ & $\varepsilon = \mathcal{H}-1$ \\ (${\mathrm{cm}}^3
\, \, {\mathrm{s}}^{-1}$) & ($\mathrm{cm}$) & ($\mathrm{cm}$) &  \\
\hline 66 & 0.415 & 0.590 & 0.42\\ \hline 91 & 0.430 & 0.619 & 0.44\\
\hline 121 & 0.445 & 0.648 & 0.46\\ \hline 148 & 0.573 & 0.856 &
0.49\\ \hline 152 & 0.590 & 0.890 & 0.51\\ \hline 162 & 0.621 & 0.940
& 0.51\\ \hline\hline
\end{tabular}
\end{center}
\caption{The magnitude of the jump $\varepsilon$, at different volumetric
flow rates $Q$.}
\label{tab1}
\end{table}

In the experiment we have also seen that the flow becomes turbulent 
after the jump, with the formation of vortices. We have not made any 
measurement for this region as we have been interested in the laminar 
flow only. However, we have noticed that for low volumetric flow rates,
turbulence in the subcritical flow region dies down appreciably, as
the flow proceeds downstream. On the other hand, for high flow rates,
turbulence is seen to be sustained right upto the outer boundary of
the flow. Qualitatively this is what it should be. Turbulent 
fluctuations derive their energy from the mean flow. If the mean flow 
is more energetic, such as it should be for higher flow rates, then 
turbulent effects will linger in the flow that much longer. Regarding 
this issue, it should be possible 
for us, at least in an order-of-magnitude sense, to make a theoretical 
estimate of the Reynolds number of the turbulent flow, immediately 
after the jump. In the shallow layer flow, the largest possible 
turbulent eddies should have a characteristic length scale that should 
at most be of the order of the flow height $h$, in the subcritical 
region. The characteristic turnover velocity of the eddies should 
likewise be of the order of (but less than) the velocity of surface 
gravity waves, $(gh)^{1/2}$. The Reynolds number of the flow, 
$\mathcal{R}_\mathrm{e}$, should therefore be given by 
$\mathcal{R}_\mathrm{e} \sim \nu^{-1}(gh^3)^{1/2}$.
In our experiment, typical values of the flow constants $g$ and $\nu$
would be $1000 \, \, \mathrm{cm \, \, sec^{-2}}$ and 
$10^{-2} \, \, \mathrm{cm^2 \, \, sec^{-1}}$, respectively. From 
Table \ref{tab1}, for various values of $Q$, we may have an estimate 
of the characteristic values of the flow height $h$ immediately after 
the jump. These measurements should then typically give 
$\mathcal{R}_\mathrm{e} \sim 1000$, 
an estimate, whose direct verification, however, would be beyond the scope 
of our experiment. 

\section*{Acknowledgements}
SBS is grateful for discussions with Prof. Deepak Dhar. 
AKR thanks Dr. Tapas K. Das for drawing attention to
some recent works in the subject of analogue gravity.

\end{document}